\documentclass[%
 aip,
 adv,%
 amsmath,amssymb,
reprint,%
groupedaddress,
onecolumn,
author-numerical,%
]{revtex4-1}

\usepackage{pifont}
\usepackage{graphicx}% Include figure files
\usepackage{dcolumn}% Align table columns on decimal point
\usepackage{bm}% bold math
\usepackage{etex}
\usepackage[etex=true,export]{adjustbox}
\begin{document}

%\preprint{AIP/123-QED}

\title[]{Impact of the magnetic proximity effect in Pt on the total magnetic moment of Pt/Co/Ta trilayers studied by x-ray resonant magnetic reflectivity}% Force line breaks with \\
%\thanks{Footnote to title of article.}

\author{Anastasiia Moskaltsova}
\email{amoskaltsova@physik.uni-bielefeld.de}

\author{Jan Krieft}%

\author{Dominik Graulich}

\author{Tristan Matalla-Wagner}

\author{Timo Kuschel}
\affiliation{%
Center for Spinelectronic Materials and Devices, Department of Physics, Bielefeld University, Universit{\"a}tsstra{\ss}e 25, 33615 Bielefeld, Germany}

\date{\today}

\begin{abstract}
In this work, we study the influence of the magnetic proximity effect (MPE) in Pt on the total magnetic moment of thin film trilayer systems consisting of the ferromagnet (FM) Co adjacent to the heavy metals (HMs) Pt and Ta. We investigate the trilayer systems HM$\textsubscript{1}$/FM/HM$\textsubscript{2}$ with different stacking order as well as a reference bilayer without any MPE. X-ray resonant magnetic reflectivity (XRMR) is a powerful tool to probe induced magnetism, especially when buried at interfaces in a multilayer. By using XRMR, we are able to obtain magnetic depth profiles of the structural, optical and magnetic parameters. By fitting the experimental data with a Gaussian-like magnetooptic profile taking the structural roughness at the interface into account, we can extract the magnetic moment of the spin-polarized layer. Comparing the obtained moments to the measured total moment of the sample, we can determine the impact of the MPE on the total magnetic moment of the system. Such information can be critical for analyzing spin transport experiments, including spin-orbit torque and spin Hall angle measurements, where the saturation magnetization \textit{M}$\textsubscript{s}$ has to be taken into account. Therefore, by combining magnetization measurements and XRMR methods we were able to get a complete picture of the magnetic moment distribution in these trilayer systems containing spin-polarized Pt. 
\end{abstract}

%                             % Classification Scheme.

\maketitle

\section{Introduction}

In the recent years, materials with large spin Hall angle such as Ta, Pt, W etc. have been widely used in spintronics. Various spintronic effects, such as the (inverse) spin Hall effect \cite{Dyakonov, Hirsch,Kato,Saitoh2006,Hoffmann2013,Sinova2015}, spin-pumping \cite{Tserkovnyak2002,Caminale}, the spin Seeback effect \cite{Uchida2010,Meier2013,Meier2015,Uchida2016} and spin Hall magnetoresistance \cite{Nakayama,Althammer2013,Chen2016} utilize systems consisting of heavy metal (HM)/ ferromagnet (FM). In such a system, a spin-orbit torque (SOT) \cite{Miron,Gambardella2011,Liu,Avci2017} induced by the spin Hall effect acts on the FM magnetization. If the torque is large enough, it can eventually switch the magnetization. Numerous works study SOT phenomena due to its use for the SOT-based magnetoresistive random-access memory (MRAM) \cite{MRAM,GarelloMRAM}. Such SOT-MRAM can potentially surpass the spin-transfer torque MRAM \cite{Ralph2008} by exhibiting faster switching  and requiring lower operating current densities \cite{Ahmed}. To calculate SOT switching efficiencies and spin Hall angle of the HM, it is essential to know the magnetic moment of the FM. \newline
\indent However, in HM/FM systems the magnetic proximity effect (MPE) can occur and thus contribute to the total magnetic moment. This effect arises when the HM is close to the Stoner criterion \cite{Stoner} (such as Pt) and in proximity to the FM. Thus, a thin layer of the HM becomes spin-polarized. This phenomenon has been intensively studied in the past by x-ray magnetic circular  dichroism (XMCD) \cite{Schütz1989,Schütz1990,Antel1999,Wilhelm,Manna2014}. However, x-ray resonant magnetic reflectivity (XRMR)\cite{remagx} has been proven to be more advantageous over XMCD in magnetic depth profile studies. Moreover, it was shown by XRMR that the MPE is independent from the Pt thickness \cite{TimoPRL} and FM thickness, as well \cite{TimoIEEE}. In addition, the strength of the MPE increases linearly with the FM moment \citep{Klewe,Polina2018} and can be manipulated by the use of a Ta buffer layer \cite{Mukhopadhyay2019}.\newline
\indent The influence of the MPE on SOT efficiencies has been recently studied for bilayer systems such as Pt/Co$\textsubscript2$FeAl \cite{Peterson2018}, Pt/Co and Au$\textsubscript{0.25}$Pt$\textsubscript{0.75}$/Co \cite{Zhu2018}. Peterson \textit{et al.} \cite{Peterson2018} attribute the anomalous increase in magnetoresistance at low temperature to the MPE. The authors further show that this contribution suppresses the field-like SOT efficiency by nearly a factor of 4 at 20 K compared to room temperature. At the same time Zhu \textit{et al.} \cite{Zhu2018} discuss irrelevance of the MPE in the bilayer systems with thin Co layer. Here, the authors compare the magnetic moments of the systems obtained via vibrating sample magnetometry (VSM) to the \textit{M}$\textsubscript{s}$ obtained from XMCD measurements on Pt/Co multilayers. As a result, the authors show that annealing enhances the MPE in both cases (Pt/Co and Au$\textsubscript{0.25}$Pt$\textsubscript{0.75}$/Co). However, Zhu and colleagues then claim that the correlation between the MPE and SOTs magnitudes is minimal and can be neglected. Hence, the MPE role in spin transport experiments and its contribution to the total magnetic moment is still an open question. \newline
\indent In this paper, we take advantage of combining XRMR and VSM, with VSM serving as an additional supportive technique, to investigate the correlation between induced magnetic moment in Pt due to MPE and total magnetic moment of the system. After addition of the extracted Pt magnetic moment to the literature Co values, we compare the result with the VSM-measured total magnetic moment of the samples and get excellent agreement. This demonstrates that the spin-polarized Pt should be taken into account when analyzing spin transport data using the total magnetic moment of the sample. 
\section{Experiment}

We study here two types of trilayers of opposite stacking order: sub//Ta 6/Co 2.5/Pt 6/cap and sub//Pt 6/Co 2.5/Ta 6/cap (nominal thickness in nm). In addition, we analyze the reference bilayer sub//Ta 6/Co 2.5/cap. The substrate (sub) for all samples is Si/SiO$\textsubscript{2}$ 50. The capping layer (cap) is also the same for all the samples and consists of MgO 2/ TaO$\textsubscript{x}$, where the top TaO$\textsubscript{x}$ layer is obtained by natural oxidation of 2 nm thin metallic Ta. The samples were deposited at room temperature by DC magnetron sputter deposition, apart from MgO which was obtained by RF sputtering. The total magnetic moment of the samples was measured utilizing a 7 T Cryogen Free Magnet system with VSM at room temperature and in-plane applied magnetic field. XRMR was used to probe the structural and magnetic depth profile, study the MPE of the Pt layer and calculate the induced Pt magnetic moment. \newline
\indent XRMR measurements were performed at room temperature at the resonant scattering and diffraction beamline P09 \cite{P09} at the third generation PETRA III synchrotron facility at DESY (Hamburg, Germany). We measured x-ray resonant reflectivity (XRR) curves in $\theta$ - 2$\theta$ scattering geometry using circularly polarized x-rays with resonant (11566.5 eV) photon energy corresponding to the Pt \textit{L}$\textsubscript{3}$ absorption edge. An external magnetic field of $\pm$90 mT was applied in the scattering plane parallel to the sample surface and the reflected intensity was detected. The magnetic field direction was alternated during the scans, while the polarity of x-rays was kept fixed. The degree of circular polarization of the x-rays was 99\%. We then define the non-magnetic reflectivity as \textit{I} = $\frac{(I\textsubscript{+}+I\textsubscript{-})}{2}$, where $I\textsubscript{+}$ and $I\textsubscript{-}$ correspond to the XRR intensity at positive and negative magnetic fields, respectively. The XRMR asymmetry ratio is defined as $\Delta I = \frac{(I\textsubscript{+}-I\textsubscript{-})}{(I\textsubscript{+}+I\textsubscript{-})}$. \newline
\indent The obtained XRR curves and asymmetry ratios were fit using the ReMagX software \cite{remagx}, partially following the fitting strategy described by Klewe \textit{et al.} \cite{Klewe}. Unlike Klewe and co-authors, we perform all the measurements at resonant conditions, thus structural parameters have to be fitted in addition to the optical constants. The XRR data was fit using the recursive Paratt algorithm \cite{Paratt}. The complex refractive index of a material is defined as \textit{n} = 1 - $\delta$ +  \textit{i}$\beta$ with the real part $\delta$ as dispersion and the imaginary part $\beta$ as absorption, connected via the Kramers - Kronig relation. We use here the layer mode, where the stack is represented as a list of layers corresponding to different compounds.  The roughness is modeled with the matrix formalism, utilizing the layer segmentation approximation of the ReMagX tool \cite{remagxweb}. The extracted structural parameters were then used for the asymmetry ratio fitting. \newline
\indent We model the asymmetry ratio based on the Zak matrix formalism \cite{Zak}, where the additional magnetic contribution is simulated by a convolution of a Gaussian spin polarization depth profile with the interface roughness at the Co/Pt interface resulting in a Gaussian-like magnetic Pt depth profile. The goodness of fit ($\chi^2$) is defined as the sum of the squared error, logarithmically weighted for XRR and linearly for the XRMR. It represents the differences between the experimental and simulated reflectivity curves and asymmetry ratios. The maximum value of $\Delta\beta$ and the effective thickness of the spin-polarized layer, defined as full width at half maximum (FWHM) of the Gaussian-like profile, were then extracted. Important to note that $\Delta\delta$ is expected to be zero throughout the whole stack at the resonant energy of the Pt \textit{L}$\textsubscript{3}$ absorption edge as derived from \textit{ab initio} calculations \cite{TimoPRL} which are also used to relate the $\Delta\beta$ depth profile, obtained by XRMR, to the final magnetic moment values per Pt atom.

\section{Results}
\begin{figure}[h]
\adjincludegraphics[max width=\linewidth]{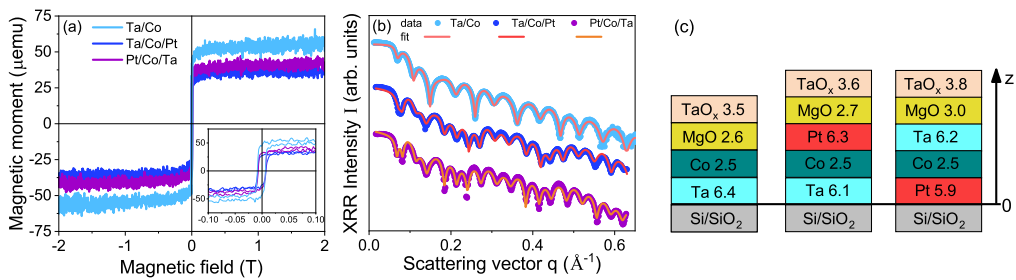}
\caption{(a) Magnetization curves measured via VSM. Inset shows a close-up of the small magnetic field region. (b) XRR data and fits for the three samples. (c) Stacks representation  with the structural parameters obtained from the XRR fits (thickness in nm). Z-axis is defined vertically with 0 being located at the substrate/stack interface. 
\label{Fig1}}
\end{figure}

We first characterize the samples by VSM, where we extract saturation fields and saturated magnetic moments.  The corresponding VSM curves are shown in Fig.~\ref{Fig1}(a). We then compare the obtained moments with the literature value for Co which is 1.7 $\mu\textsubscript{B}$ per Co atom according to the Slater-Pauling curve \cite{chikazumi} (see Tab. \ref{Table1}). For the Ta/Co bilayer the measured magnetic moment is close to the literature one and thus $\mu\textsubscript{VSM} = \mu\textsubscript{Co,lit}$. However, when comparing the trilayers containing Pt in proximity to Co, the measured magnetic moment is higher than expected for Co. We thus conclude that for the trilayer samples the MPE contributes to the total magnetic moment. 
 
\begin{figure*}[h]
\adjincludegraphics[max width=\linewidth]{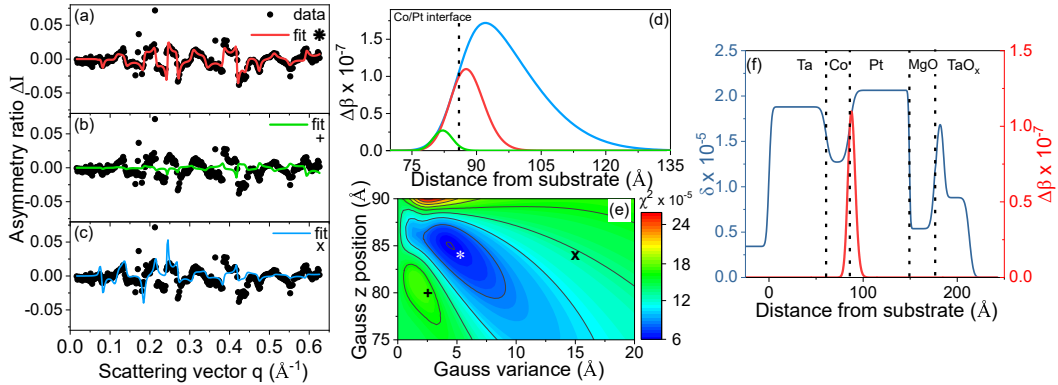}
\caption{(a), (b), (c) Asymmetry ratio and various fits for the  Ta/Co/Pt trilayer. (d) Corresponding magnetooptic Pt depth profiles from the various asymmetry ratio fits. (e) $\chi^2$ map plot of the Gauss variance vs. the Gauss z position. The \textbf{+}, \ding{107} and \textbf{$\times$} symbols identify the respective asymmetry ratios. (f) Optic $\delta$ depth profile (solid blue line) and best fit magnetooptic $\Delta\beta$ depth profile (solid red line), the same as in (d).  \label{Fig2}}
\end{figure*}

We use the XRR curves (Fig.~\ref{Fig1}(b)) to extract the structural parameters of the studied stacks. For the trilayers we fit here the resonant XRR curves obtained at a wavelength $\lambda$ = 1.07192 \AA~ (11566.5 eV), while for the Ta/Co bilayer off-resonant XRR was obtained by Philips X'Pert Pro MPD x-ray diffractometer at wavelength $\lambda$ = 1.5406 \AA~ (8047.7 eV), corresponding to Cu K\textsubscript{$\alpha$} energy. The extracted thicknesses of the stacks are shown in Fig.~\ref{Fig1}(c). \newline 
 \begin{table*}
\caption{Co literature values, Pt induced magnetic moments and magnetic moments measured by VSM.}\label{Table1}
\resizebox{\textwidth}{!} {%
\setlength{\tabcolsep}{6pt}
\begin{tabular}{@{}llllll@{}}
     \toprule  %\vspace{-6pt}\\
	 	 	 
    Sample & A~$(\textrm{cm}^2)$ & $\mu\textsubscript{Co,lit}$ \cite{chikazumi} ($\mu$emu) &\multicolumn{1}{c|}{$\mu\textsubscript{Pt,XRMR}$ ($\mu$emu)} & $\mu\textsubscript{Co,lit}$ + $\mu\textsubscript{Pt,XRMR}$ ($\mu$emu) & $\mu\textsubscript{VSM}$ ($\mu$emu)\\
    %\hline
    Ta/Co & 0.15~$\pm$0.04 & 53.6 & \multicolumn{1}{c|}{0.0} & 53.6 & 53.0 $\pm$4.9\\
    %\hline
    Ta/Co/Pt & 0.09~$\pm$0.03 & 32.0 & \multicolumn{1}{c|}{2.8~$\pm$1.0} & 34.8~$\pm$1.0 & 35.6 $\pm$1.1\\
    %\hline
    Pt/Co/Ta & 0.12~$\pm$0.04 & 42.8 & \multicolumn{1}{c|}{2.2~$\pm$0.9} & 45.0~$\pm$0.9 & 44.9 $\pm$1.9\\
\botrule

\end{tabular}}
\end{table*}
\indent To calculate the MPE-induced Pt magnetic moment, we take a closer look at the asymmetry ratio curves. Figure \ref{Fig2} presents the asymmetry ratios for the Ta/Co/Pt trilayer sample. In order to study the $\chi^2$ dependence on various parameters, we first perform 2D $\chi^2$ mapping \cite{JK} by varying Gaussian profile parameters, namely variance and z position, keeping the profile amplitude fixed. The obtained $\chi^2$ map is shown in Fig.~\ref{Fig2}(e). By using the calculated landscape we were able to narrow the parameters range and perform more detailed analysis of the fits. Figures~\ref{Fig2}(a), (b) and (c) show various asymmetry ratios obtained from different depth profile models, corresponding to the highlighted areas in the $\chi^2$ landscape (the \textbf{+}, \textbf{\ding{107}} and \textbf{$\times$} symbols), while Fig.~\ref{Fig2}(d) illustrates the change of the final magnetooptic $\Delta\beta$ depth profiles for the chosen models. Figure~\ref{Fig2}(a) represents the best fit with minimal goodness of fit $\chi^2$ = 5.47 $\times 10^{-5}$. If we choose the Gauss profile parameters (variance and z position) far from the $\chi^2$ minimum, the resulting asymmetry ratios do not fit the experimental data neither qualitatively, nor quantitatively (Figs. \ref{Fig2}(b) and (c)). For those cases, the convoluted $\Delta\beta$ depth profiles are not physically meaningful, due to the median being inside Co layer (Fig. \ref{Fig2}(d) green line) or far into Pt layer (Fig. \ref{Fig2}(d) light blue line).
The final best fit Gauss parameters lie inside the global $\chi^2$ minimum area. The ultimate optic and magnetooptic profiles are presented in Fig.~\ref{Fig2}(f). The magnetooptic profile (solid blue line) shows the $\delta$ depth profile along the sample stack. The top TaO$\textsubscript{x}$ layer shows in both cases a change of $\delta$ within the layer, which can be explained by different stages of oxidation of the metallic Ta. \newline
\begin{figure}[h]
\includegraphics[scale=0.6]{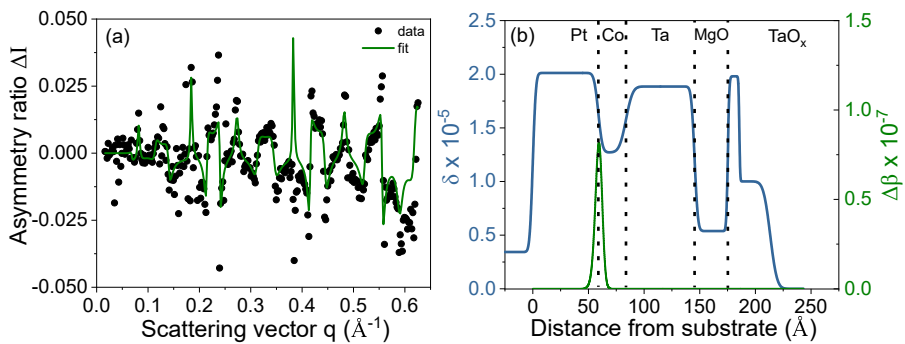}
\caption{ (a) Pt/Ta/Co sample asymmetry ratio and the corresponding fit. (b) Optical $\delta$ profile (left y-axis) and magnetooptic Gaussian-like $\Delta\beta$ depth profile obtained from the fit (right y-axis). \label{Fig3}}
\end{figure}
The similar procedure was performed for the Pt/Co/Ta system with reversed stacking order and the resulting asymmetry ratio simulation as well as the optic and magnetooptic profiles are represented in Figs.~\ref{Fig3}(a) and (b), respectively. The modeled asymmetry ratio results in a slightly higher $\chi^2$ value compared to the other stacking order ($\chi^2$ = 8.47 $\times 10^{-5}$). This might result in higher uncertainty of $\Delta\beta$ and final Pt magnetic moment. \newline
\indent By comparing the obtained $\Delta\beta$ values to the \textit{ab initio} calculations \cite{TimoPRL} we then obtain induced Pt magnetic moments of 0.56 $\mu\textsubscript{B}$ per Pt atom at the interface for Ta/Co/Pt and 0.42 $\mu\textsubscript{B}$ per atom at the interface for Pt/Co/Ta. The obtained moments are comparable with the previously reported for Pt/Fe \cite{Klewe} or Pt/Co$\textsubscript{1-x}$Fe$\textsubscript{x}$ \cite{Polina2018} bilayers. The difference in the induced moments among the trilayers might arise due to differently textured crystal growth for Pt grown on Co compared to Pt grown on the substrate Si/SiO$\textsubscript{2}$. For example, it was recently shown that different textures of Pt thin films result in a different magnitude of the MPE \cite{Mukhopadhyay2019}. \newline
\indent The extracted thicknesses of the Pt spin-polarized layer are 0.9 nm and 0.8 nm for Ta/Co/Pt and Pt/Co/Ta respectively. These thicknesses are slightly lower than the previously reported values for MPE in Pt/Fe bilayers \cite{Klewe}. The extracted information is further used to recalculate the magnetic moment in $\mu$emu and to compare with the measured moment of the sample. We find that the obtained Pt moment is 2.8 $\mu$emu for Ta/Co/Pt trilayer, while for the reversed order trilayer it is 2.2 $\mu$emu. When we add these values to the Co literature values, the resulting moments are comparable to the measured ones (see Table \ref{Table1}). Thus we conclude that the measured high magnetic moment can be represented as a sum of Co and Pt induce moments $\mu\textsubscript{VSM} = \mu\textsubscript{Co,lit} + \mu\textsubscript{Pt,XRMR}$.
\section{Conclusion}
In conclusion, we have shown that when the MPE is present in a system, it is necessary to correctly take it into account for the total magnetic moment estimation. While XRMR allowed us to probe the spin-depth profile of complex stacks, the VSM technique has been proven to aid in accessing the quantitative estimation of the induced magnetic moment of Pt. With help of the $\chi^2$ landscape mapping we were able to compare different combinations of fit parameters yielding the best fit and physically consistent spin-depth profiles. The obtained magnetic moments of 0.56 $\mu\textsubscript{B}$ and 0.42 $\mu\textsubscript{B}$ per Pt atom at the interface  are comparable to the previously reported for simple bilayers consisting of Pt and 3d magnetic metal alloys. The calculated induced magnetic moments of the spin-polarized Pt layer correlate well with the additional increased total moment for the samples containing Pt.
\section{Acknowledgments}
We acknowledge DESY (Hamburg, Germany), a member of the Helmholtz Association HGF, for the provision of experimental facilities. Parts of this research were carried out at PETRA III and we would like to thank Dr. Sonia Francoual for support in setting up the XRMR experiment at beamline P09. The authors thank G{\"u}nter Reiss for making available the laboratory equipment in Bielefeld. We further thank Tobias Pohlmann for support during the beamtime and acknowledge financial support by the Deutsche Forschungsgemeinschaft (DFG) within the individual grants RE 1052/22-2 and RE 1052/42-1. Tristan Matalla-Wagner was directly supported by Bielefeld University.
 
\bibliography{referncesfile}
\end{document}